\documentclass[pra,superscriptaddress,twocolumn,amsmath,amssymb]{revtex4-1} 
\usepackage{graphicx}
\usepackage{dcolumn}
\usepackage{bm}
\usepackage{amsmath}
\usepackage{float}

\begin{document}

\title{Efficient generation of  twin photons at telecom wavelengths with 10 GHz repetition-rate tunable comb laser}

\author{Rui-Bo Jin}
\email{ruibo@nict.go.jp}
\affiliation{National Institute of Information and Communications Technology (NICT), 4-2-1 Nukui-Kitamachi, Koganei, Tokyo 184-8795, Japan}
\author{Ryosuke Shimizu}
\affiliation{ University of Electro-Communications (UEC), Tokyo 182-8585, Japan}
\author{Isao Morohashi}
\affiliation{National Institute of Information and Communications Technology (NICT), 4-2-1 Nukui-Kitamachi, Koganei, Tokyo 184-8795, Japan}\
\author{Kentaro Wakui}
\affiliation{National Institute of Information and Communications Technology (NICT), 4-2-1 Nukui-Kitamachi, Koganei, Tokyo 184-8795, Japan}
\author{Masahiro Takeoka}
\affiliation{National Institute of Information and Communications Technology (NICT), 4-2-1 Nukui-Kitamachi, Koganei, Tokyo 184-8795, Japan}
\author{Shuro Izumi}
\affiliation{National Institute of Information and Communications Technology (NICT), 4-2-1 Nukui-Kitamachi, Koganei, Tokyo 184-8795, Japan}
\affiliation{Sophia University,7-1Kioicho,Chiyoda-ku,Tokyo 102-8554,Japan}
\author{Takahide Sakamoto}
\affiliation{National Institute of Information and Communications Technology (NICT), 4-2-1 Nukui-Kitamachi, Koganei, Tokyo 184-8795, Japan}
\author{Mikio Fujiwara}
\affiliation{National Institute of Information and Communications Technology (NICT), 4-2-1 Nukui-Kitamachi, Koganei, Tokyo 184-8795, Japan}
\author{Taro Yamashita}
\affiliation{National Institute of Information and Communications Technology (NICT), 588-2 Iwaoka, Kobe 651-2492, Japan}
\author{Shigehito Miki}
\affiliation{National Institute of Information and Communications Technology (NICT), 588-2 Iwaoka, Kobe 651-2492, Japan}
\author{Hirotaka Terai}
\affiliation{National Institute of Information and Communications Technology (NICT), 588-2 Iwaoka, Kobe 651-2492, Japan}
\author{Zhen Wang}
\affiliation{National Institute of Information and Communications Technology (NICT), 588-2 Iwaoka, Kobe 651-2492, Japan}
\affiliation{Shanghai Institute of Microsystem and Information Technology, Chinese Academy of Sciences (CAS), 865 Changning Road, Shanghai 200050, China}
\author{Masahide Sasaki}
\affiliation{National Institute of Information and Communications Technology (NICT), 4-2-1 Nukui-Kitamachi, Koganei, Tokyo 184-8795, Japan}

\date{\today }

\begin{abstract}
Efficient generation and detection of indistinguishable twin photons are at the core of quantum information and communications technology (Q-ICT).
These photons are conventionally generated by spontaneous parametric down conversion (SPDC), which is a probabilistic process, and hence occurs at a limited rate, which restricts wider applications of Q-ICT.
To increase the rate, one had to excite SPDC by higher pump power, while it inevitably produced more unwanted multi-photon components, harmfully degrading quantum interference visibility.
Here we solve this problem by using recently developed 10 GHz repetition-rate-tunable comb laser \cite{Sakamoto2008, Morohashi2012}, combined with a group-velocity-matched nonlinear crystal \cite{Jin2013OE, Jin2014OE}, and superconducting nanowire single photon detectors \cite{Miki2013, Yamashita2013}.
They operate at telecom wavelengths more efficiently with less noises than conventional schemes, those typically operate at visible and near infrared wavelengths generated by a 76 MHz Ti Sapphire laser and detected by Si detectors
\cite{Lu2007, Huang2011, Yao2012}.
We could show high interference visibilities, which are free from the pump-power induced degradation.
Our laser, nonlinear crystal, and detectors constitute a powerful tool box, which will pave a way to implementing quantum photonics circuits with variety of good and low-cost telecom components, and will eventually realize scalable Q-ICT in optical infra-structures.

\end{abstract}

\maketitle

\section{Introduction}

Since the first experimental realization of quantum teleportation \cite{Bouwmeester1997},
many experiments with multiphoton entanglement have been demonstrated \cite{Lu2007,Pan2012},
and currently expanded to eight photons, employing multiple SPDC crystals \cite{Huang2011,Yao2012}.
In order to increase the scale of entanglement further, the generation probability per SPDC crystal
must be drastically improved without degrading the quantum indistinguishability of photons.
Unfortunately, however, a dilemma always exists in SPDC: higher pump power is required for higher generation probability,
while it degrades quantum interference visibility due to unwanted multi-pair emissions, leading to the increase of error rates
in entanglement-based quantum key distribution (QKD) \cite{Fujiwara2014}
and photonic quantum information processing \cite{Knill2001}.

When $n$ SPDC sources are used, the 2$n$-fold coincidence counts (CC) can be estimated as
\begin{equation}
\label{eq1}
CC=f p^n \eta^{2n},
\end{equation}
where $f$ is the repetition rate of the pump laser,
$p$ is the generation probability of one photon-pair per pulse,
$\eta$ is the overall efficiency,
which is the product of the collecting efficiency of the whole optical system and the detecting efficiency of the detectors.
The $p$ should be restricted to less than 0.1, so that the effect of unwanted multi-pair emissions can be negligible.
So the pump power is tuned for $p \le 0.1$.

The value of $p$ is not high in the conventional photon source.
A standard technology is based on SPDC at visible and near infrared wavelengths using a BBO crystal pumped by the second harmonic of the femto-second laser pulses
from a Ti Sapphire laser, whose repetition rate is 76 MHz \cite{Lu2007, Huang2011, Yao2012}.
In this case, the probability $p$ had not been able to go beyond 0.06, because the second harmonic power was limited to 300-900\,mW for a fundamental laser power of 1-3\,W.
Therefore, recent efforts have been focused on increasing the pump power \cite{Krischek2010}.

Recently the periodically poled KTP (PPKTP) crystals attract much attention because it can achieve $p\sim$0.1 (0.6) at telecom wavelengths
with a  pump power of 80 (400) mW thanks to the quasi-phase matching (QPM) technique \cite{Jin2013SNSPD}.
When waveguide structure is employed, $p$ can be $10^3$ times higher than the bulk crystal \cite{Tanzilli2001, Zhong2012}.
Unfortunately, however, the constraint $p \le 0.1$ should be met in these cases too.
The $\eta$ is already maximized by careful alignment in laboratories,  e.g., the typical $\eta$ value is about 0.2-0.3 \cite{Huang2011, Yao2012, Jin2013SNSPD}.
Thus $p$ and $\eta$ have almost reach their maxima.
The remaining effective way is to improve the repetition rate of the pump laser, $f$.

In this work, we demonstrate a novel photon source pumped by a recently developed repetition-rate-tunable comb laser in a range of 10-0.625 GHz \cite{Sakamoto2008, Morohashi2012}.
This laser can operate in relatively low pulse energy, while keeping high average power, thanks to a high repetition rate.
The low pulse energy would result in the reduction of the multiple-pair emission.
At the same time a high counting rate would be expected owing to the high average power.
The SPDC based on a group-velocity-matched PPKTP (GVM-PPKTP) crystal can achieve very high spectral purity of the constituent photons \cite{Jin2013OE, Jin2014OE}.
Furthermore, the photons are detected by the state-of-the-art superconducting nanowire single photon detectors (SNSPDs) \cite{Miki2013, Yamashita2013},
which have a much higher efficiency than that of traditional InGaAs avalanche photodiode (APD).
%


\section{Experiment}

\begin{figure*}[tbp]
\centering
\includegraphics[width= 0.90 \textwidth]{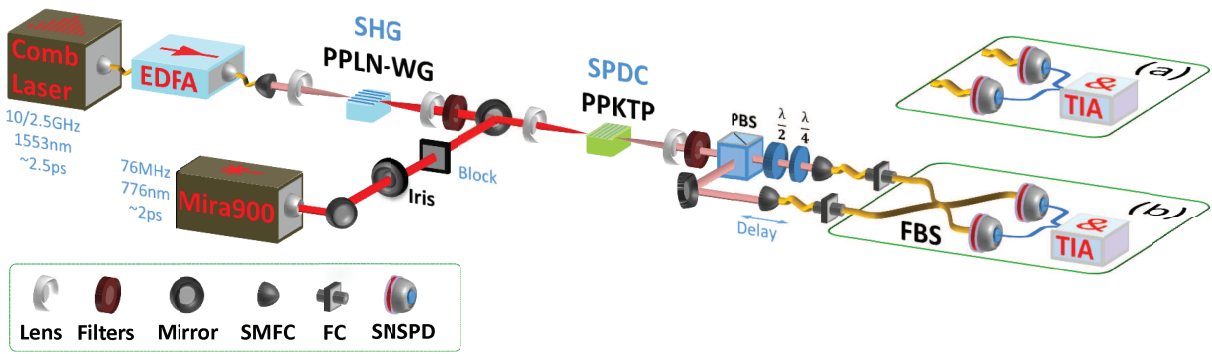}
\caption{ \textbf{ The Experimental setup.}
A comb laser with a 10-0.625 GHz tunable-repetition-rate  at 1553 nm wavelength is amplified by a high-power erbium-doped fiber amplifier (EDFA), and  frequency doubled by a 10-mm-long type-0 periodically poled lithium niobate wave guide (PPLN-WG, by NTT Electronics, polling period=18.2 $\mu$m).
Then, after filtered by short-pass filters, the 776.5 nm  laser light from second harmonic generation (SHG)  pumps a 30-mm  type-II  periodically poled potassium titanyl phosphate crystal (PPKTP, by Raicol, poling period= 46.1 $\mu$m, temperature = $47.8\,^{\circ}\mathrm{C}$) for SPDC.
The downconverted photons, the signal and idler, are filtered by long-pass filters, separated by a polarization beamsplitter (PBS), and  coupled into single mode fibers (SMF) by couplers (SMFC).
(a) The collected photons are directly detected by SNSPDs, which are connected to a time interval analyzer (TIA) for the measurement of  Time of Arrival (ToA).
(b) SMFs are connected to a fiber beamsplitter (FBS) by two fiber connectors (FC),  for the test of  Hong-Ou-Mandel dip. A half-wave plate and a quarter-wave plate are added to guarantee the signal and idler photons have the same polarization in FBS.
For comparison, we  also  introduce a 76 MHz pump laser (Mira900, at 776.5 nm, around 2 ps). An iris is inserted for 76 MHz laser to make the beam waist comparable with the comb laser.
 }
\label{setup}
\end{figure*}

The experimental setup is shown in Fig.\,\ref{setup}.
%
%
%
%
%
%
%
%
The picosecond pulses from the comb laser are generated with the following principles \cite{Sakamoto2008, Morohashi2012}.
A continuous-wave (cw) light emitted from a single-mode laser diode (LD) with a wavelength of around 1553 nm
is led into a Mach-Zehnder-modulator (MZM) and is converted to a comb signal with 10 GHz in spacing and 300 GHz in bandwidth.
The MZM is fabricated on a LiNbO$_3$ crystal and is driven by a 10 GHz radio-frequency signal.
Because the comb signal has linear chirp, it can be formed to a picosecond pulse train with a repetition rate of 10 GHz by chirp compensation using a single-mode fiber.
The comb laser also includes a pulse picker, so that the repetition frequency of the pulse train can be changed in the range of 10-0.625 GHz.
In this experiment, we  keep the temporal width around 2.5 ps.
Fig.\,\ref{comblaser}(a) shows the spectrum of this laser at  2.5 GHz repetition rate.
See the Appendix for more spectral and temporal information of this comb laser.
For more details of this kind of comb lasers, see Refs. \cite{Sakamoto2007, Morohashi2008}.
%
%
%
\begin{figure*}[tbp]
\centering
\includegraphics[width= 0.98 \textwidth]{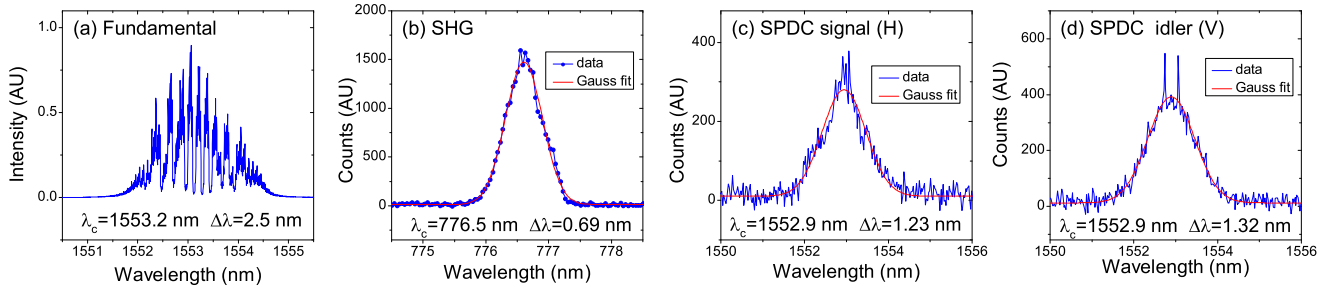}
\caption{  \textbf{ The spectra.}
          (a) The spectra of the comb laser at 2.5 GHz repetition rates.
          (b) A typical spectrum of SHG, pumped by the fundamental laser.
          (c, d) Spectra of the signal (horizontal polarization)
              and idler (vertical polarization) photons generated in SPDC.
         } \label{comblaser}
\end{figure*}
%

Generating a high-power second harmonic  light (SHG) is a key point in this experiment.
Since the average power per pulse of the comb laser is very low, we choose a periodically poled lithium niobate wave guide (PPLN-WG) for SHG.
We tested both  10 GHz and 2.5 GHz repetition rate lasers.
We found the SHG power with 2.5 GHz repetition rate was more stable than that with 10 GHz repetition rate.
Therefore, the data in this experiment are mainly obtained by using 2.5 GHz repetition rate.
With the input 2.5 GHz repetition rate fundamental laser at a power of 500 mW, we obtained 42 mW SHG power.
After filtered by several short-pass filters to cut the fundamental light, we finally achieved a net SHG power of 35 mW.
The transmission loss of the PPLN-WG was around 50\%.
Fig.\,\ref{comblaser}(b) is the  spectrum of 776.5nm SHG laser, measured by a spectrometer (SpectraPro-2300i, Acton Research Corp.).
Interestingly, it can be noticed that the comb structure no-longer exists in the SHG spectrum, which may be caused by a sum-frequency-generation process.

For SPDC, the nonlinear crystal used in  this experiment is  a PPKTP crystal, which satisfy the GVM condition at telecom wavelength \cite{Jin2013OE, Konig2004, Evans2010, Gerrits2011, Eckstein2011}.
Thanks to the GVM condition, the spectral purity is much higher at telecom wavelength than that at visible wavelengths \cite{Jin2014OE}.
This spectrally pure photon source is very useful for multi-photon interference between independent sources \cite{Mosley2008a, Jin2011, Jin2013PRA}.
Figure\,\ref{comblaser}(c, d) are the observed spectra of the signal and idler photons,  measured by a spectrometer (SpctraPro-2500i, Acton Research Corp.).
The FWHMs of the twin photons are about 1.2-1.3 nm, similar as the spectral width of the photons pumped by 76 MHz laser \cite{Jin2013OE}.

Our superconducting nanowire single photon detectors (SNSPDs) have a system detection efficiency (SDE) of around 70\% with a dark count rate (DCR) less than 1 kcps \cite{Miki2013, Yamashita2013, Jin2013SNSPD}.
The SNSPD also has a wide spectral response range that covers at least from 1470 nm to 1630 nm wavelengths \cite{Jin2013SNSPD}.
The measured timing jitter and  dead time (recovery time) were  68 ps \cite{Miki2013} and 40 ns \cite{Miki2007}.

The performance with the 2.5 GHz source is evaluated in terms of a signal to noise ratio (SNR) \cite{Broome2011} and a Hong-Ou-Mandel (HOM) interference \cite{Hong1987}, in comparison with the 76 MHz laser.
The SNR test is carried out in the detecting configuration of Fig.\,\ref{setup}(a), while the HOM test is performed in that of Fig.\,\ref{setup}(b).

\section{Result 1:  Signal to noise ratio test}

\begin{figure*}[t]
\centering
\includegraphics[width= 0.98 \textwidth]{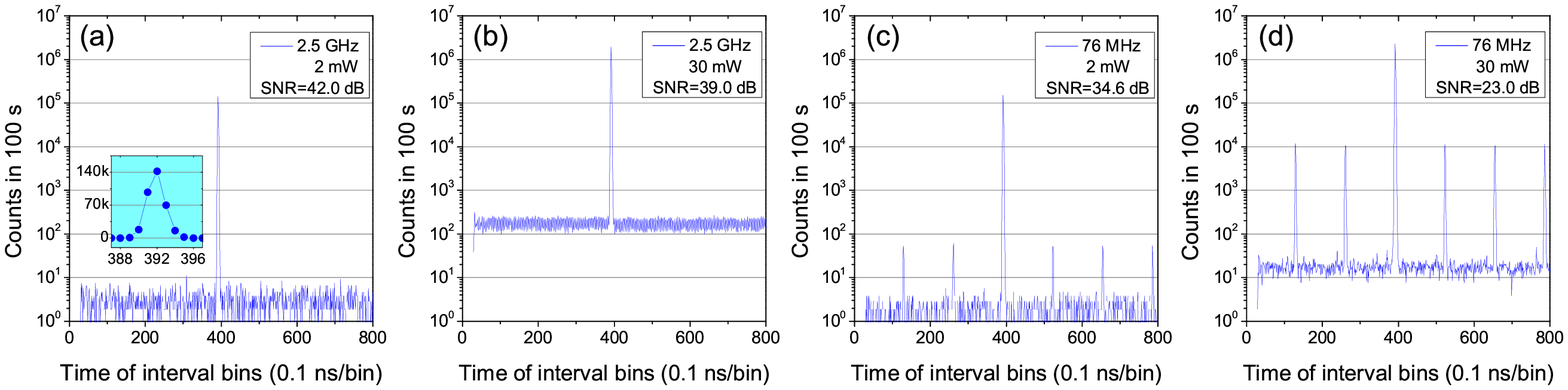}
\caption{
 \textbf{  Time of Arrival (ToA) data.}
The data in each figure was accumulated in 100 seconds  for 2.5 GHz and 76 MHz lasers at 2 mW and 30 mW.
The peak intervals are 0.4 ns in (a, b) and 13.1 ns in (c, d).
The measured SNR values are 42.0 dB, 39.0 dB, 34.6 dB and 23.0 dB for (a-d), respectively.
Inset in (a) is an enlarged view of the main peak in a linear scale.
 } \label{TIA}
\end{figure*}

The SNR is defined as the ratio of single pair emission rate over the double pair emission rate \cite{Broome2011}.
These rates can be evaluated by Time of Arrival (ToA) data, which are shown in Fig.\,\ref{TIA}(a-d).
%
%
%
%
Each data consists of the main peak and side peaks, and we define the peak counts as the value of each peak.
The side peaks are not visible in Fig.\,\ref{TIA}(a, b) because the resolution of the detector system ($\sim$0.5 ns, as seen in the inset in (a)) is comparable to the pulse interval of 2.5 GHz laser (0.4 ns).  So, we set the averaged maximal counts as the side peak values.
The side peaks are recorded when a second SPDC occurs (in the stop channel) conditioned on a first SPDC (in the start channel) occurs at the main peaks.
Therefore, the side peaks correspond to the rate of 2-pair components in SPDC, while the main peaks correspond to the rate of 1-pair plus 2-pair components in SPDC.
So the SNR can be calculated in dB as
\begin{equation}
\label{eq2}
SNR=10 \log_{10}{[(\text{main peak} -\text{side peak})/\text{side peak}]}.
\end{equation}
The measured SNR are 42.0 dB, 39.0 dB, 34.6 dB, 23.0 dB for Fig.\,\ref{TIA}(a-d), respectively.

Theoretical calculation unveils that the SNR is proportional to the inverse of average photon numbers per pulse at a lower pump power \cite{Broome2011}.
This claim can be experimentally  verified by comparing Fig.\,\ref{TIA}(c) and (d) with the 76 MHz laser.
When the pump power increases from 2 mW to 30 mW, the SNR is decreased by 34.6 - 23.0 = 11.6 dB.
It agrees well with the 30 mW / 2 mW = 15 times (11.7 dB) increase of average photon number per pulse.

Next, we compare the result in  Fig.\,\ref{TIA}(b) and (d),  so as to confirm the validity of the definition for  side peak values in  Fig.\,\ref{TIA}(a) and (b).
At 30 mW pump power, the coincidence counts are 48 kcps and 56 kcps for 2.5 GHz and 76 MHz laser, respectively, as seen from  Fig.\,\ref{TIA}(b) and (d).
Then the average photon numbers per pulse are estimated to be 0.00021 and 0.0079, correspondingly.
The average photon pair per pulse for the 2.5 GHz laser is 0.0079 / 0.00021 = 37.6 times (15.8 dB) lower than that of the 76 MHz laser.
Recall the SNR difference between 2.5 GHz and 76 MHz of 39.0 - 23.0 = 16.0 dB.
This consistency verifies the validity of the definition for  side peak values in  Fig.\,\ref{TIA}(a) and (b).

Finally, we estimate the SNR values for the case of the 2.5 GHz laser at 2 mW in Fig.\,\ref{TIA}(a).
The SNR at 2 mW should ideally increase by 11.7 dB (15 times), from 39.0 dB (30 mW in Fig.\,\ref{TIA}(b)) to 50.7 dB.
Actually, however, the measured SNR is only 42.0 dB.
This discrepancy is mainly due to the dark counts by the detectors and the accidental counts by stray photons.

\section{Result 2:  Hong-Ou-Mandel interference test}

\begin{figure*}[tbp]
\centering
\includegraphics[width= 0.85 \textwidth]{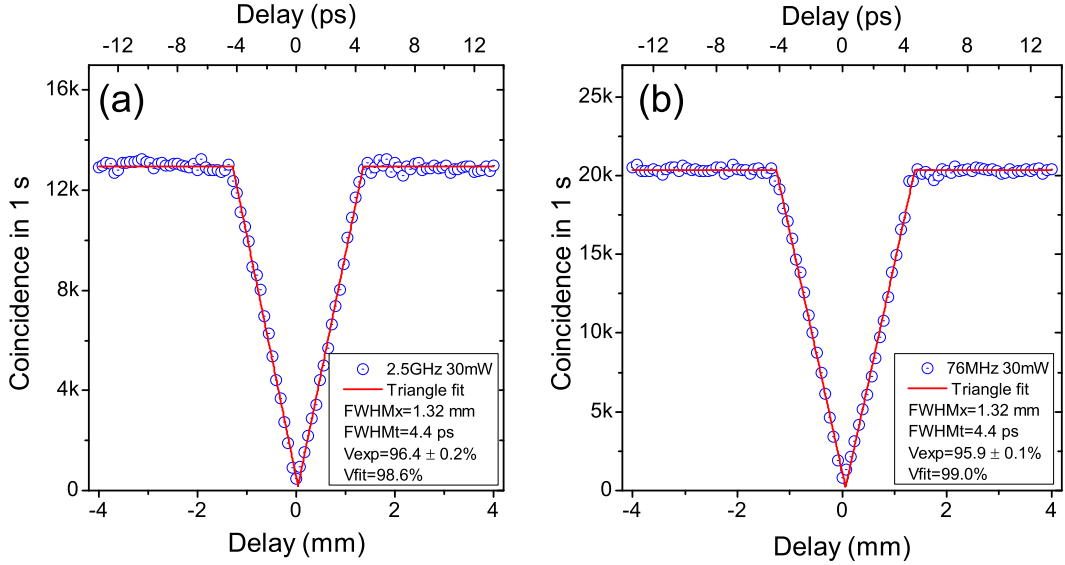}
\caption{
\textbf{  Hong-Ou-Mandel (HOM) dip.}
HOM dip for 2.5 GHz (a) and 76 MHz (b) repetition rate lasers with pump power of 30 mW, fitted with triangle shape function.}
\label{HOMI}
\end{figure*}

We then carried out the HOM interference test to evaluate the performance of a twin photon source.
We firstly worked with 30 mW pump power for 2.5 GHz and  76 MHz repetition rate lasers, and achieved raw visibilities of  96.4 $\pm$ 0.2\% and 95.9 $\pm$ 0.1\%, respectively, as shown in Fig.\,\ref{HOMI}(a, b).
%
%
%
%
The triangle-shape of the HOM dip in  Fig.\,\ref{HOMI}(a, b) is caused by
the group-velocity matching condition in PPKTP crystal at telecom wavelengths \cite{Kuzucu2005, Shimizu2009}.
The widths of the dips in Fig.\,\ref{HOMI} (a, b) are similar, around 1.33 mm (4.4 ps), since the width of the dip is determined by the length of the crystal \cite{Ansari2014}.
The high visibilities in Fig.\,\ref{HOMI} confirmed the high indistinguishability of the generated photons.

\begin{figure*}[tbp]
\centering
\includegraphics[width=0.5 \textwidth]{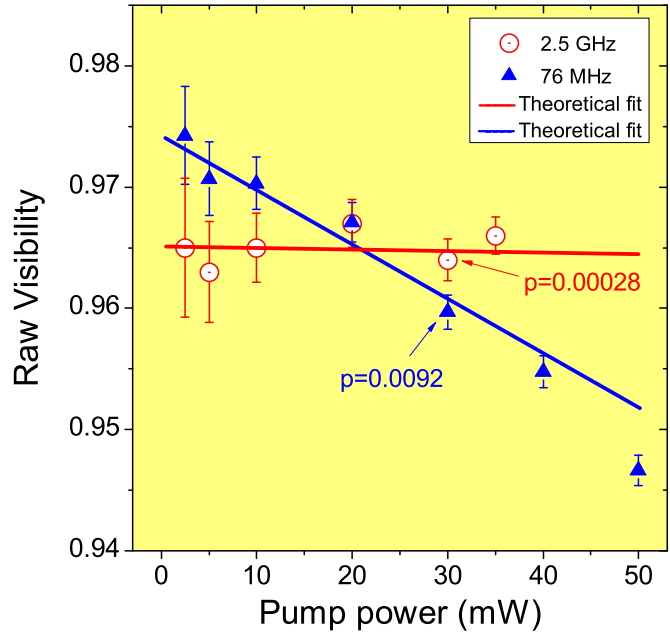}
\caption{\textbf{Raw visibilities  versus pump powers for   2.5 GHz and 76 MHz   lasers.}
At 30 mW, the average photon numbers per pulse were 0.00028 and 0.0092 for   2.5 GHz and 76 MHz   lasers.
The uncertainties of these visibilities were derived using Poissonian errors on the coincidence counts.
The  solid lines are the theoretical fit. } \label{VsV}
\end{figure*}

To compare the different performance of the 2.5 GHz and 76 MHz lasers, we repeated the HOM interference test  at different pump powers.
Without subtract any background counts, we compare the raw visibilities of HOM dip  at different pump powers in  Fig.\,\ref{VsV}.
%
%
%
%
At a low pump power of 2.5 mW, the 76 MHz laser has a visibility of 97.4 $\pm$ 0.4\%,  slightly higher than the result by the 2.5 GHz laser, 96.5 $\pm$0.6\%.
At 30 mW, the average photon numbers per pulse were 0.00028 and 0.0092 for   2.5 GHz and 76 MHz   lasers.
Note the average photon numbers per pulse in  the HOM interference test were slightly higher than that in the ToA test, because we slightly improved the experimental condition in the test of HOM interference.

It is noteworthy that the visibilities by the 76 MHz laser decrease rapidly when the pump powers increase.
In contrast, the visibilities by the 2.5 GHz laser  shows  almost no decrease up to 35 mW, the maximum SHG power we have achieved in experiment.
To fit this  experimental data, we construct a theoretical model.
In this model, the transmittance loss of the signal and idler photons are effectively described by two beam splitters,  and the mode-matching efficiency $\eta_M$ between the signal and idler are also represented by two beam splitters with the transmittance of $\eta_M$.
The transmittance loss is obtained from the experimental condition, while the mode-matching efficiency are searched so as to fit the experimental data.
The numerical analyses suggest that the HOM visibility is extremely sensitive to the mode matching efficiency, $\eta_M$.
However, it is not easy to estimate the experimental $\eta_M$ value with enough accuracy.
In  Fig.\,\ref{VsV},  the data are fitted with $\eta_M$ values of 0.9828 and 0.9878 for 2.5 GHz laser and 76 MHz laser, respectively.
Higher $\eta_M$  value for the 76 MHz laser is reasonable, because the indistinguishablites of the twin photons generated by the 76 MHz laser is slightly better than  that by the 2.5 GHz laser,
which can be roughly checked by the spectra of the signal and idler in  Fig.\,\ref{comblaser}(c) and (d).
After the transmittance efficiency and the mode-matching efficiency are fixed, the HOM visibilities are only determined by $p$, the average photon pairs per pulse (i.e., the generation probability of one pair per pulse).
The low value $p$ by the 2.5 GHz laser guarantees its high visibilities at high pump powers.
In Fig.\,\ref{VsV}, the theoretical model fitted well with our experimental data.
See more details of our model in the Appendix.

\section{Discussion and Outlook}
With the theoretical model (See the Appendix), we further calculate the visibilities at high pump powers,  as shown in Tab. \ref{table3W}.
It is interesting to note that,  at  high pump power up to 3 W, the visibility by 76 MHz laser will decrease to 62.4\%, while the 2.5 GHz laser still can keep  the visibility higher than 90 \%, mainly thanks to the low average photon numbers per pulse.
To experimentally demonstrate this high visibilities at high pump powers in the future,  we could update the PPKTP bulk crystal to a PPKTP waveguide.
Also, SHG power of the comb laser at 10 GHz repetition rate  needs to be improved.
See  the Appendix for the HOM interference of this comb laser at 10 GHz repetition rate.
\begin{table}
\centering
\begin{tabular}{cccc}
\hline \hline
           &30 mW           &$\sim$300 mW       &$\sim$3 W    \\
76  MHz    &0.0092 (96.1\%)   &0.092 (86.1\%)     &0.92 (62.4\%) \\
2.5 GHz    &0.00028 (96.4\%)   &0.0028 (96.0\%)   &0.028 (91.8\%)  \\
\hline \hline
\end{tabular}
\caption{\label{table3W} \textbf{Visibilities at high pump powers.} The estimated average photon pairs per pulse and the corresponding visibilites at different pump powers for 76  MHz and 2.5 GHz laser.  The visibilities are directly determined by the average photon pairs per pulse.}
\end{table}
Nevertheless, our experimental results in  Fig.\,\ref{VsV} have clearly shown the non-degradation of HOM visibilites at high pump powers.

We notice that   many  other schemes have been reported to  reduce the multi-pair emission.
Broome \emph{et al}  demonstrated to reduce the multi-pair emission  by  temporally multiplexing the pulsed pump lasers by two times \cite{Broome2011}.
Ma \emph{et al} tried to reduce such effects by  multiplexing four independent SPDC sources \cite{Ma2011}.
All the previous methods have a limited effect, because the units they multiplexed were limited.
If they increase the multiplexed units, the setup will be very complex.
The GHz-repetition-rate-laser pumped  photon source in our scheme provide a very simple and effective way to reduce the multi-pair emission.
In addition, GHz repetition rate lasers are now commercially available
\cite{Morris2014, GigaOptics, M2laser}.

Therefore, our scheme will be a reasonable option to construct the next generation of twin photon sources with high brightness, low multi-pair emission and high detection efficiency.
In the traditional twin photon source technologies, 76 MHZ pump laser is compatible with a BBO crystal (with maximum $p \sim 0.06$, corresponding to photon pair generation rate of 5 MHz) and Si avalanche photodiode (with acceptable maximal photon numbers of 5-10 MHz).
In the next generation of twin photon sources, the 10 GHz laser should be combined with a high efficiency crystals, e.g., PPKTP crystal (or waveguide, with maximum  $p\sim 0.1$, corresponding to photon pair generation rate of 1 GHz),
and high speed detectors, e.g., the SNSPD (with acceptable maximal photon numbers of 25-100 MHz).
Consequently, we expect more than tenfold brighter photon source in conjunction with both low multiple photon pairs production and high spectral purity.
Further, a repetition tunability allow us to obtain an optimal generation probability in a pulse without sacrificing a photon counting rate.

\section{Conclusion}
We have demonstrated a  twin photon source pumped by a 10-GHz-repetition-rate tunable comb laser.
The photons are generated from GVM-PPKTP crystal and detected by highly efficient SNSPDs.
The SNR test and HOM interference test with 2.5 GHz laser showed a high SNR and high visibilities not degrading at high pump powers,
much higher than  that pumped by the 76 MHz laser.
The high-repetition-rate pump laser, the GVM-PPKTP crystal, and the highly efficient detectors constitute a powerful tool box  at the telecom wavelengths.
We believe our tool box  may have a variety of applications in the future quantum infromantion and communiction technologies.

\section*{Acknowledgements}
The authors thank  N. Singh for insightful discussion.
This work was supported by the Founding Program for World-Leading Innovative R\&D on Science and Technology (FIRST).


\begin{thebibliography}{36}%
\makeatletter
\providecommand \@ifxundefined [1]{%
 \@ifx{#1\undefined}
}%
\providecommand \@ifnum [1]{%
 \ifnum #1\expandafter \@firstoftwo
 \else \expandafter \@secondoftwo
 \fi
}%
\providecommand \@ifx [1]{%
 \ifx #1\expandafter \@firstoftwo
 \else \expandafter \@secondoftwo
 \fi
}%
\providecommand \natexlab [1]{#1}%
\providecommand \enquote  [1]{``#1''}%
\providecommand \bibnamefont  [1]{#1}%
\providecommand \bibfnamefont [1]{#1}%
\providecommand \citenamefont [1]{#1}%
\providecommand \href@noop [0]{\@secondoftwo}%
\providecommand \href [0]{\begingroup \@sanitize@url \@href}%
\providecommand \@href[1]{\@@startlink{#1}\@@href}%
\providecommand \@@href[1]{\endgroup#1\@@endlink}%
\providecommand \@sanitize@url [0]{\catcode `\\12\catcode `\$12\catcode
  `\&12\catcode `\#12\catcode `\^12\catcode `\_12\catcode `\%12\relax}%
\providecommand \@@startlink[1]{}%
\providecommand \@@endlink[0]{}%
\providecommand \url  [0]{\begingroup\@sanitize@url \@url }%
\providecommand \@url [1]{\endgroup\@href {#1}{\urlprefix }}%
\providecommand \urlprefix  [0]{URL }%
\providecommand \Eprint [0]{\href }%
\providecommand \doibase [0]{http://dx.doi.org/}%
\providecommand \selectlanguage [0]{\@gobble}%
\providecommand \bibinfo  [0]{\@secondoftwo}%
\providecommand \bibfield  [0]{\@secondoftwo}%
\providecommand \translation [1]{[#1]}%
\providecommand \BibitemOpen [0]{}%
\providecommand \bibitemStop [0]{}%
\providecommand \bibitemNoStop [0]{.\EOS\space}%
\providecommand \EOS [0]{\spacefactor3000\relax}%
\providecommand \BibitemShut  [1]{\csname bibitem#1\endcsname}%
\let\auto@bib@innerbib\@empty
\bibitem [{\citenamefont {Sakamoto}\ \emph {et~al.}(2008)\citenamefont
  {Sakamoto}, \citenamefont {Kawanishi},\ and\ \citenamefont
  {Tsuchiya}}]{Sakamoto2008}%
  \BibitemOpen
  \bibfield  {author} {\bibinfo {author} {\bibfnamefont {T.}~\bibnamefont
  {Sakamoto}}, \bibinfo {author} {\bibfnamefont {T.}~\bibnamefont {Kawanishi}},
  \ and\ \bibinfo {author} {\bibfnamefont {M.}~\bibnamefont {Tsuchiya}},\
  }\href {\doibase 10.1364/OL.33.000890} {\bibfield  {journal} {\bibinfo
  {journal} {Opt. Lett.}\ }\textbf {\bibinfo {volume} {33}},\ \bibinfo {pages}
  {890} (\bibinfo {year} {2008})}\BibitemShut {NoStop}%
\bibitem [{\citenamefont {Morohashi}\ \emph {et~al.}(2012)\citenamefont
  {Morohashi}, \citenamefont {Oikawa}, \citenamefont {Tamura}, \citenamefont
  {Aoki}, \citenamefont {Sakamoto}, \citenamefont {Kawanishi},\ and\
  \citenamefont {Hosako}}]{Morohashi2012}%
  \BibitemOpen
  \bibfield  {author} {\bibinfo {author} {\bibfnamefont {I.}~\bibnamefont
  {Morohashi}}, \bibinfo {author} {\bibfnamefont {M.}~\bibnamefont {Oikawa}},
  \bibinfo {author} {\bibfnamefont {Y.}~\bibnamefont {Tamura}}, \bibinfo
  {author} {\bibfnamefont {S.}~\bibnamefont {Aoki}}, \bibinfo {author}
  {\bibfnamefont {T.}~\bibnamefont {Sakamoto}}, \bibinfo {author}
  {\bibfnamefont {T.}~\bibnamefont {Kawanishi}}, \ and\ \bibinfo {author}
  {\bibfnamefont {I.}~\bibnamefont {Hosako}},\ }in\ \href {\doibase
  10.1364/CLEO_SI.2012.CF1N.7} {\emph {\bibinfo {booktitle} {Conference on
  Lasers and Electro-Optics 2012}}}\ (\bibinfo  {publisher} {Optical Society of
  America},\ \bibinfo {year} {2012})\ p.\ \bibinfo {pages} {CF1N.7}\BibitemShut
  {NoStop}%
\bibitem [{\citenamefont {Jin}\ \emph {et~al.}(2013{\natexlab{a}})\citenamefont
  {Jin}, \citenamefont {Shimizu}, \citenamefont {Wakui}, \citenamefont
  {Benichi},\ and\ \citenamefont {Sasaki}}]{Jin2013OE}%
  \BibitemOpen
  \bibfield  {author} {\bibinfo {author} {\bibfnamefont {R.-B.}\ \bibnamefont
  {Jin}}, \bibinfo {author} {\bibfnamefont {R.}~\bibnamefont {Shimizu}},
  \bibinfo {author} {\bibfnamefont {K.}~\bibnamefont {Wakui}}, \bibinfo
  {author} {\bibfnamefont {H.}~\bibnamefont {Benichi}}, \ and\ \bibinfo
  {author} {\bibfnamefont {M.}~\bibnamefont {Sasaki}},\ }\href@noop {}
  {\bibfield  {journal} {\bibinfo  {journal} {Opt. Express}\ }\textbf {\bibinfo
  {volume} {21}},\ \bibinfo {pages} {10659} (\bibinfo {year}
  {2013}{\natexlab{a}})}\BibitemShut {NoStop}%
\bibitem [{\citenamefont {Jin}\ \emph {et~al.}(2014)\citenamefont {Jin},
  \citenamefont {Shimizu}, \citenamefont {Wakui}, \citenamefont {Fujiwara},
  \citenamefont {Yamashita}, \citenamefont {Miki}, \citenamefont {Terai},
  \citenamefont {Wang},\ and\ \citenamefont {Sasaki}}]{Jin2014OE}%
  \BibitemOpen
  \bibfield  {author} {\bibinfo {author} {\bibfnamefont {R.-B.}\ \bibnamefont
  {Jin}}, \bibinfo {author} {\bibfnamefont {R.}~\bibnamefont {Shimizu}},
  \bibinfo {author} {\bibfnamefont {K.}~\bibnamefont {Wakui}}, \bibinfo
  {author} {\bibfnamefont {M.}~\bibnamefont {Fujiwara}}, \bibinfo {author}
  {\bibfnamefont {T.}~\bibnamefont {Yamashita}}, \bibinfo {author}
  {\bibfnamefont {S.}~\bibnamefont {Miki}}, \bibinfo {author} {\bibfnamefont
  {H.}~\bibnamefont {Terai}}, \bibinfo {author} {\bibfnamefont
  {Z.}~\bibnamefont {Wang}}, \ and\ \bibinfo {author} {\bibfnamefont
  {M.}~\bibnamefont {Sasaki}},\ }\bibfield  {booktitle} {\emph {\bibinfo
  {booktitle} {Optics Express}},\ }\href@noop {} {\bibfield  {journal}
  {\bibinfo  {journal} {Opt. Express}\ }\textbf {\bibinfo {volume} {22}},\
  \bibinfo {pages} {11498} (\bibinfo {year} {2014})}\BibitemShut {NoStop}%
\bibitem [{\citenamefont {Miki}\ \emph {et~al.}(2013)\citenamefont {Miki},
  \citenamefont {Yamashita}, \citenamefont {Terai},\ and\ \citenamefont
  {Wang}}]{Miki2013}%
  \BibitemOpen
  \bibfield  {author} {\bibinfo {author} {\bibfnamefont {S.}~\bibnamefont
  {Miki}}, \bibinfo {author} {\bibfnamefont {T.}~\bibnamefont {Yamashita}},
  \bibinfo {author} {\bibfnamefont {H.}~\bibnamefont {Terai}}, \ and\ \bibinfo
  {author} {\bibfnamefont {Z.}~\bibnamefont {Wang}},\ }\href@noop {} {\bibfield
   {journal} {\bibinfo  {journal} {Opt. Express}\ }\textbf {\bibinfo {volume}
  {21}},\ \bibinfo {pages} {10208} (\bibinfo {year} {2013})}\BibitemShut
  {NoStop}%
\bibitem [{\citenamefont {Yamashita}\ \emph {et~al.}(2013)\citenamefont
  {Yamashita}, \citenamefont {Miki}, \citenamefont {Terai},\ and\ \citenamefont
  {Wang}}]{Yamashita2013}%
  \BibitemOpen
  \bibfield  {author} {\bibinfo {author} {\bibfnamefont {T.}~\bibnamefont
  {Yamashita}}, \bibinfo {author} {\bibfnamefont {S.}~\bibnamefont {Miki}},
  \bibinfo {author} {\bibfnamefont {H.}~\bibnamefont {Terai}}, \ and\ \bibinfo
  {author} {\bibfnamefont {Z.}~\bibnamefont {Wang}},\ }\bibfield  {booktitle}
  {\emph {\bibinfo {booktitle} {Optics Express}},\ }\href@noop {} {\bibfield
  {journal} {\bibinfo  {journal} {Opt. Express}\ }\textbf {\bibinfo {volume}
  {21}},\ \bibinfo {pages} {27177} (\bibinfo {year} {2013})}\BibitemShut
  {NoStop}%
\bibitem [{\citenamefont {Lu}\ \emph {et~al.}(2007)\citenamefont {Lu},
  \citenamefont {Zhou}, \citenamefont {Guhne}, \citenamefont {Gao},
  \citenamefont {Zhang}, \citenamefont {Yuan}, \citenamefont {Goebel},
  \citenamefont {Yang},\ and\ \citenamefont {Pan}}]{Lu2007}%
  \BibitemOpen
  \bibfield  {author} {\bibinfo {author} {\bibfnamefont {C.-Y.}\ \bibnamefont
  {Lu}}, \bibinfo {author} {\bibfnamefont {X.-Q.}\ \bibnamefont {Zhou}},
  \bibinfo {author} {\bibfnamefont {O.}~\bibnamefont {Guhne}}, \bibinfo
  {author} {\bibfnamefont {W.-B.}\ \bibnamefont {Gao}}, \bibinfo {author}
  {\bibfnamefont {J.}~\bibnamefont {Zhang}}, \bibinfo {author} {\bibfnamefont
  {Z.-S.}\ \bibnamefont {Yuan}}, \bibinfo {author} {\bibfnamefont
  {A.}~\bibnamefont {Goebel}}, \bibinfo {author} {\bibfnamefont
  {T.}~\bibnamefont {Yang}}, \ and\ \bibinfo {author} {\bibfnamefont {J.-W.}\
  \bibnamefont {Pan}},\ }\href@noop {} {\bibfield  {journal} {\bibinfo
  {journal} {Nat. Phys.}\ }\textbf {\bibinfo {volume} {3}},\ \bibinfo {pages}
  {91} (\bibinfo {year} {2007})}\BibitemShut {NoStop}%
\bibitem [{\citenamefont {Huang}\ \emph {et~al.}(2011)\citenamefont {Huang},
  \citenamefont {Liu}, \citenamefont {Peng}, \citenamefont {Li}, \citenamefont
  {Li}, \citenamefont {Li},\ and\ \citenamefont {Guo}}]{Huang2011}%
  \BibitemOpen
  \bibfield  {author} {\bibinfo {author} {\bibfnamefont {Y.-F.}\ \bibnamefont
  {Huang}}, \bibinfo {author} {\bibfnamefont {B.-H.}\ \bibnamefont {Liu}},
  \bibinfo {author} {\bibfnamefont {L.}~\bibnamefont {Peng}}, \bibinfo {author}
  {\bibfnamefont {Y.-H.}\ \bibnamefont {Li}}, \bibinfo {author} {\bibfnamefont
  {L.}~\bibnamefont {Li}}, \bibinfo {author} {\bibfnamefont {C.-F.}\
  \bibnamefont {Li}}, \ and\ \bibinfo {author} {\bibfnamefont {G.-C.}\
  \bibnamefont {Guo}},\ }\href@noop {} {\bibfield  {journal} {\bibinfo
  {journal} {Nat. Commun.}\ }\textbf {\bibinfo {volume} {2}},\ \bibinfo {pages}
  {546(1} (\bibinfo {year} {2011})}\BibitemShut {NoStop}%
\bibitem [{\citenamefont {Yao}\ \emph {et~al.}(2012)\citenamefont {Yao},
  \citenamefont {Wang}, \citenamefont {Xu}, \citenamefont {Lu}, \citenamefont
  {Pan}, \citenamefont {Bao}, \citenamefont {Peng}, \citenamefont {Lu},
  \citenamefont {Chen},\ and\ \citenamefont {Pan}}]{Yao2012}%
  \BibitemOpen
  \bibfield  {author} {\bibinfo {author} {\bibfnamefont {X.-C.}\ \bibnamefont
  {Yao}}, \bibinfo {author} {\bibfnamefont {T.-X.}\ \bibnamefont {Wang}},
  \bibinfo {author} {\bibfnamefont {P.}~\bibnamefont {Xu}}, \bibinfo {author}
  {\bibfnamefont {H.}~\bibnamefont {Lu}}, \bibinfo {author} {\bibfnamefont
  {G.-S.}\ \bibnamefont {Pan}}, \bibinfo {author} {\bibfnamefont {X.-H.}\
  \bibnamefont {Bao}}, \bibinfo {author} {\bibfnamefont {C.-Z.}\ \bibnamefont
  {Peng}}, \bibinfo {author} {\bibfnamefont {C.-Y.}\ \bibnamefont {Lu}},
  \bibinfo {author} {\bibfnamefont {Y.-A.}\ \bibnamefont {Chen}}, \ and\
  \bibinfo {author} {\bibfnamefont {J.-W.}\ \bibnamefont {Pan}},\ }\href@noop
  {} {\bibfield  {journal} {\bibinfo  {journal} {Nat. Photon.}\ }\textbf
  {\bibinfo {volume} {6}},\ \bibinfo {pages} {225} (\bibinfo {year}
  {2012})}\BibitemShut {NoStop}%
\bibitem [{\citenamefont {Bouwmeester}\ \emph {et~al.}(1997)\citenamefont
  {Bouwmeester}, \citenamefont {Pan}, \citenamefont {Mattle}, \citenamefont
  {Eibl}, \citenamefont {Weinfurter},\ and\ \citenamefont
  {Zeilinger}}]{Bouwmeester1997}%
  \BibitemOpen
  \bibfield  {author} {\bibinfo {author} {\bibfnamefont {D.}~\bibnamefont
  {Bouwmeester}}, \bibinfo {author} {\bibfnamefont {J.-W.}\ \bibnamefont
  {Pan}}, \bibinfo {author} {\bibfnamefont {K.}~\bibnamefont {Mattle}},
  \bibinfo {author} {\bibfnamefont {M.}~\bibnamefont {Eibl}}, \bibinfo {author}
  {\bibfnamefont {H.}~\bibnamefont {Weinfurter}}, \ and\ \bibinfo {author}
  {\bibfnamefont {A.}~\bibnamefont {Zeilinger}},\ }\href@noop {} {\bibfield
  {journal} {\bibinfo  {journal} {Nature}\ }\textbf {\bibinfo {volume} {390}},\
  \bibinfo {pages} {575} (\bibinfo {year} {1997})}\BibitemShut {NoStop}%
\bibitem [{\citenamefont {Pan}\ \emph {et~al.}(2012)\citenamefont {Pan},
  \citenamefont {Chen}, \citenamefont {Lu}, \citenamefont {Weinfurter},
  \citenamefont {Zeilinger},\ and\ \citenamefont {{\.Z}ukowski}}]{Pan2012}%
  \BibitemOpen
  \bibfield  {author} {\bibinfo {author} {\bibfnamefont {J.-W.}\ \bibnamefont
  {Pan}}, \bibinfo {author} {\bibfnamefont {Z.-B.}\ \bibnamefont {Chen}},
  \bibinfo {author} {\bibfnamefont {C.-Y.}\ \bibnamefont {Lu}}, \bibinfo
  {author} {\bibfnamefont {H.}~\bibnamefont {Weinfurter}}, \bibinfo {author}
  {\bibfnamefont {A.}~\bibnamefont {Zeilinger}}, \ and\ \bibinfo {author}
  {\bibfnamefont {M.}~\bibnamefont {{\.Z}ukowski}},\ }\href@noop {} {\bibfield
  {journal} {\bibinfo  {journal} {Rev. Mod. Phys.}\ }\textbf {\bibinfo {volume}
  {84}},\ \bibinfo {pages} {777} (\bibinfo {year} {2012})}\BibitemShut
  {NoStop}%
\bibitem [{\citenamefont {Fujiwara}\ \emph {et~al.}(2014)\citenamefont
  {Fujiwara}, \citenamefont {Yoshino}, \citenamefont {Nambu}, \citenamefont
  {Yamashita}, \citenamefont {Miki}, \citenamefont {Terai}, \citenamefont
  {Wang}, \citenamefont {Toyoshima}, \citenamefont {Tomita},\ and\
  \citenamefont {Sasaki}}]{Fujiwara2014}%
  \BibitemOpen
  \bibfield  {author} {\bibinfo {author} {\bibfnamefont {M.}~\bibnamefont
  {Fujiwara}}, \bibinfo {author} {\bibfnamefont {K.-i.}\ \bibnamefont
  {Yoshino}}, \bibinfo {author} {\bibfnamefont {Y.}~\bibnamefont {Nambu}},
  \bibinfo {author} {\bibfnamefont {T.}~\bibnamefont {Yamashita}}, \bibinfo
  {author} {\bibfnamefont {S.}~\bibnamefont {Miki}}, \bibinfo {author}
  {\bibfnamefont {H.}~\bibnamefont {Terai}}, \bibinfo {author} {\bibfnamefont
  {Z.}~\bibnamefont {Wang}}, \bibinfo {author} {\bibfnamefont {M.}~\bibnamefont
  {Toyoshima}}, \bibinfo {author} {\bibfnamefont {A.}~\bibnamefont {Tomita}}, \
  and\ \bibinfo {author} {\bibfnamefont {M.}~\bibnamefont {Sasaki}},\
  }\bibfield  {booktitle} {\emph {\bibinfo {booktitle} {Optics Express}},\
  }\href@noop {} {\bibfield  {journal} {\bibinfo  {journal} {Opt. Express}\
  }\textbf {\bibinfo {volume} {22}},\ \bibinfo {pages} {13616} (\bibinfo {year}
  {2014})}\BibitemShut {NoStop}%
\bibitem [{\citenamefont {Knill}\ \emph {et~al.}(2001)\citenamefont {Knill},
  \citenamefont {Laflamme},\ and\ \citenamefont {Milburn}}]{Knill2001}%
  \BibitemOpen
  \bibfield  {author} {\bibinfo {author} {\bibfnamefont {E.}~\bibnamefont
  {Knill}}, \bibinfo {author} {\bibfnamefont {R.}~\bibnamefont {Laflamme}}, \
  and\ \bibinfo {author} {\bibfnamefont {G.~J.}\ \bibnamefont {Milburn}},\
  }\href@noop {} {\bibfield  {journal} {\bibinfo  {journal} {Nature}\ }\textbf
  {\bibinfo {volume} {409}},\ \bibinfo {pages} {46} (\bibinfo {year}
  {2001})}\BibitemShut {NoStop}%
\bibitem [{\citenamefont {Krischek}\ \emph {et~al.}(2010)\citenamefont
  {Krischek}, \citenamefont {Wieczorek}, \citenamefont {Ozawa}, \citenamefont
  {Kiesel}, \citenamefont {Michelberger}, \citenamefont {Udem},\ and\
  \citenamefont {Weinfurter}}]{Krischek2010}%
  \BibitemOpen
  \bibfield  {author} {\bibinfo {author} {\bibfnamefont {R.}~\bibnamefont
  {Krischek}}, \bibinfo {author} {\bibfnamefont {W.}~\bibnamefont {Wieczorek}},
  \bibinfo {author} {\bibfnamefont {A.}~\bibnamefont {Ozawa}}, \bibinfo
  {author} {\bibfnamefont {N.}~\bibnamefont {Kiesel}}, \bibinfo {author}
  {\bibfnamefont {P.}~\bibnamefont {Michelberger}}, \bibinfo {author}
  {\bibfnamefont {T.}~\bibnamefont {Udem}}, \ and\ \bibinfo {author}
  {\bibfnamefont {H.}~\bibnamefont {Weinfurter}},\ }\href@noop {} {\bibfield
  {journal} {\bibinfo  {journal} {Nat. Photon.}\ }\textbf {\bibinfo {volume}
  {4}},\ \bibinfo {pages} {170} (\bibinfo {year} {2010})}\BibitemShut {NoStop}%
\bibitem [{\citenamefont {Jin}\ \emph {et~al.}(2013{\natexlab{b}})\citenamefont
  {Jin}, \citenamefont {Fujiwara}, \citenamefont {Yamashita}, \citenamefont
  {Miki}, \citenamefont {Terai}, \citenamefont {Wang}, \citenamefont {Wakui},
  \citenamefont {Shimizu},\ and\ \citenamefont {Sasaki}}]{Jin2013SNSPD}%
  \BibitemOpen
  \bibfield  {author} {\bibinfo {author} {\bibfnamefont {R.-B.}\ \bibnamefont
  {Jin}}, \bibinfo {author} {\bibfnamefont {M.}~\bibnamefont {Fujiwara}},
  \bibinfo {author} {\bibfnamefont {T.}~\bibnamefont {Yamashita}}, \bibinfo
  {author} {\bibfnamefont {S.}~\bibnamefont {Miki}}, \bibinfo {author}
  {\bibfnamefont {H.}~\bibnamefont {Terai}}, \bibinfo {author} {\bibfnamefont
  {Z.}~\bibnamefont {Wang}}, \bibinfo {author} {\bibfnamefont {K.}~\bibnamefont
  {Wakui}}, \bibinfo {author} {\bibfnamefont {R.}~\bibnamefont {Shimizu}}, \
  and\ \bibinfo {author} {\bibfnamefont {M.}~\bibnamefont {Sasaki}},\
  }\href@noop {} {\bibfield  {journal} {\bibinfo  {journal} {arXiv:1309.1221}\
  } (\bibinfo {year} {2013}{\natexlab{b}})}\BibitemShut {NoStop}%
\bibitem [{\citenamefont {Tanzilli}\ \emph {et~al.}(2001)\citenamefont
  {Tanzilli}, \citenamefont {de~Riedmatten}, \citenamefont {Tittel},
  \citenamefont {Zbinden}, \citenamefont {Baldi}, \citenamefont {De~Micheli},
  \citenamefont {Ostrowsky},\ and\ \citenamefont {Gisin}}]{Tanzilli2001}%
  \BibitemOpen
  \bibfield  {author} {\bibinfo {author} {\bibfnamefont {S.}~\bibnamefont
  {Tanzilli}}, \bibinfo {author} {\bibfnamefont {H.}~\bibnamefont
  {de~Riedmatten}}, \bibinfo {author} {\bibfnamefont {H.}~\bibnamefont
  {Tittel}}, \bibinfo {author} {\bibfnamefont {H.}~\bibnamefont {Zbinden}},
  \bibinfo {author} {\bibfnamefont {P.}~\bibnamefont {Baldi}}, \bibinfo
  {author} {\bibfnamefont {M.}~\bibnamefont {De~Micheli}}, \bibinfo {author}
  {\bibfnamefont {D.~B.}\ \bibnamefont {Ostrowsky}}, \ and\ \bibinfo {author}
  {\bibfnamefont {N.}~\bibnamefont {Gisin}},\ }\bibfield  {booktitle} {\emph
  {\bibinfo {booktitle} {Electronics Letters}},\ }\href@noop {} {\bibfield
  {journal} {\bibinfo  {journal} {Electron. Lett.}\ }\textbf {\bibinfo {volume}
  {37}},\ \bibinfo {pages} {26} (\bibinfo {year} {2001})}\BibitemShut {NoStop}%
\bibitem [{\citenamefont {Zhong}\ \emph {et~al.}(2012)\citenamefont {Zhong},
  \citenamefont {Wong}, \citenamefont {Restelli},\ and\ \citenamefont
  {Bienfang}}]{Zhong2012}%
  \BibitemOpen
  \bibfield  {author} {\bibinfo {author} {\bibfnamefont {T.}~\bibnamefont
  {Zhong}}, \bibinfo {author} {\bibfnamefont {F.~N.~C.}\ \bibnamefont {Wong}},
  \bibinfo {author} {\bibfnamefont {A.}~\bibnamefont {Restelli}}, \ and\
  \bibinfo {author} {\bibfnamefont {J.~C.}\ \bibnamefont {Bienfang}},\
  }\href@noop {} {\bibfield  {journal} {\bibinfo  {journal} {Opt. Express}\
  }\textbf {\bibinfo {volume} {20}},\ \bibinfo {pages} {26868} (\bibinfo {year}
  {2012})}\BibitemShut {NoStop}%
\bibitem [{\citenamefont {Sakamoto}\ \emph {et~al.}(2007)\citenamefont
  {Sakamoto}, \citenamefont {Kawanishi},\ and\ \citenamefont
  {Izutsu}}]{Sakamoto2007}%
  \BibitemOpen
  \bibfield  {author} {\bibinfo {author} {\bibfnamefont {T.}~\bibnamefont
  {Sakamoto}}, \bibinfo {author} {\bibfnamefont {T.}~\bibnamefont {Kawanishi}},
  \ and\ \bibinfo {author} {\bibfnamefont {M.}~\bibnamefont {Izutsu}},\ }\href
  {\doibase 10.1364/OL.32.001515} {\bibfield  {journal} {\bibinfo  {journal}
  {Opt. Lett.}\ }\textbf {\bibinfo {volume} {32}},\ \bibinfo {pages} {1515}
  (\bibinfo {year} {2007})}\BibitemShut {NoStop}%
\bibitem [{\citenamefont {Morohashi}\ \emph {et~al.}(2008)\citenamefont
  {Morohashi}, \citenamefont {Sakamoto}, \citenamefont {Sotobayashi},
  \citenamefont {Kawanishi}, \citenamefont {Hosako},\ and\ \citenamefont
  {Tsuchiya}}]{Morohashi2008}%
  \BibitemOpen
  \bibfield  {author} {\bibinfo {author} {\bibfnamefont {I.}~\bibnamefont
  {Morohashi}}, \bibinfo {author} {\bibfnamefont {T.}~\bibnamefont {Sakamoto}},
  \bibinfo {author} {\bibfnamefont {H.}~\bibnamefont {Sotobayashi}}, \bibinfo
  {author} {\bibfnamefont {T.}~\bibnamefont {Kawanishi}}, \bibinfo {author}
  {\bibfnamefont {I.}~\bibnamefont {Hosako}}, \ and\ \bibinfo {author}
  {\bibfnamefont {M.}~\bibnamefont {Tsuchiya}},\ }\href@noop {} {\bibfield
  {journal} {\bibinfo  {journal} {Opt. Lett.}\ }\textbf {\bibinfo {volume}
  {33}},\ \bibinfo {pages} {1192} (\bibinfo {year} {2008})}\BibitemShut
  {NoStop}%
\bibitem [{\citenamefont {K{\"o}nig}\ and\ \citenamefont
  {Wong}(2004)}]{Konig2004}%
  \BibitemOpen
  \bibfield  {author} {\bibinfo {author} {\bibfnamefont {F.}~\bibnamefont
  {K{\"o}nig}}\ and\ \bibinfo {author} {\bibfnamefont {F.~N.~C.}\ \bibnamefont
  {Wong}},\ }\href@noop {} {\bibfield  {journal} {\bibinfo  {journal} {Appl.
  Phys. Lett.}\ }\textbf {\bibinfo {volume} {84}},\ \bibinfo {pages} {1644}
  (\bibinfo {year} {2004})}\BibitemShut {NoStop}%
\bibitem [{\citenamefont {Evans}\ \emph {et~al.}(2010)\citenamefont {Evans},
  \citenamefont {Bennink}, \citenamefont {Grice}, \citenamefont {Humble},\ and\
  \citenamefont {Schaake}}]{Evans2010}%
  \BibitemOpen
  \bibfield  {author} {\bibinfo {author} {\bibfnamefont {P.~G.}\ \bibnamefont
  {Evans}}, \bibinfo {author} {\bibfnamefont {R.~S.}\ \bibnamefont {Bennink}},
  \bibinfo {author} {\bibfnamefont {W.~P.}\ \bibnamefont {Grice}}, \bibinfo
  {author} {\bibfnamefont {T.~S.}\ \bibnamefont {Humble}}, \ and\ \bibinfo
  {author} {\bibfnamefont {J.}~\bibnamefont {Schaake}},\ }\href@noop {}
  {\bibfield  {journal} {\bibinfo  {journal} {Phys. Rev. Lett.}\ }\textbf
  {\bibinfo {volume} {105}},\ \bibinfo {pages} {253601} (\bibinfo {year}
  {2010})}\BibitemShut {NoStop}%
\bibitem [{\citenamefont {Gerrits}\ \emph {et~al.}(2011)\citenamefont
  {Gerrits}, \citenamefont {Stevens}, \citenamefont {Baek}, \citenamefont
  {Calkins}, \citenamefont {Lita}, \citenamefont {Glancy}, \citenamefont
  {Knill}, \citenamefont {Nam}, \citenamefont {Mirin}, \citenamefont
  {Hadfield}, \citenamefont {Bennink}, \citenamefont {Grice}, \citenamefont
  {Dorenbos}, \citenamefont {Zijlstra}, \citenamefont {Klapwijk},\ and\
  \citenamefont {Zwiller}}]{Gerrits2011}%
  \BibitemOpen
  \bibfield  {author} {\bibinfo {author} {\bibfnamefont {T.}~\bibnamefont
  {Gerrits}}, \bibinfo {author} {\bibfnamefont {M.~J.}\ \bibnamefont
  {Stevens}}, \bibinfo {author} {\bibfnamefont {B.}~\bibnamefont {Baek}},
  \bibinfo {author} {\bibfnamefont {B.}~\bibnamefont {Calkins}}, \bibinfo
  {author} {\bibfnamefont {A.}~\bibnamefont {Lita}}, \bibinfo {author}
  {\bibfnamefont {S.}~\bibnamefont {Glancy}}, \bibinfo {author} {\bibfnamefont
  {E.}~\bibnamefont {Knill}}, \bibinfo {author} {\bibfnamefont {S.~W.}\
  \bibnamefont {Nam}}, \bibinfo {author} {\bibfnamefont {R.~P.}\ \bibnamefont
  {Mirin}}, \bibinfo {author} {\bibfnamefont {R.~H.}\ \bibnamefont {Hadfield}},
  \bibinfo {author} {\bibfnamefont {R.~S.}\ \bibnamefont {Bennink}}, \bibinfo
  {author} {\bibfnamefont {W.~P.}\ \bibnamefont {Grice}}, \bibinfo {author}
  {\bibfnamefont {S.}~\bibnamefont {Dorenbos}}, \bibinfo {author}
  {\bibfnamefont {T.}~\bibnamefont {Zijlstra}}, \bibinfo {author}
  {\bibfnamefont {T.}~\bibnamefont {Klapwijk}}, \ and\ \bibinfo {author}
  {\bibfnamefont {V.}~\bibnamefont {Zwiller}},\ }\href@noop {} {\bibfield
  {journal} {\bibinfo  {journal} {Opt. Express}\ }\textbf {\bibinfo {volume}
  {19}},\ \bibinfo {pages} {24434} (\bibinfo {year} {2011})}\BibitemShut
  {NoStop}%
\bibitem [{\citenamefont {Eckstein}\ \emph {et~al.}(2011)\citenamefont
  {Eckstein}, \citenamefont {Christ}, \citenamefont {Mosley},\ and\
  \citenamefont {Silberhorn}}]{Eckstein2011}%
  \BibitemOpen
  \bibfield  {author} {\bibinfo {author} {\bibfnamefont {A.}~\bibnamefont
  {Eckstein}}, \bibinfo {author} {\bibfnamefont {A.}~\bibnamefont {Christ}},
  \bibinfo {author} {\bibfnamefont {P.~J.}\ \bibnamefont {Mosley}}, \ and\
  \bibinfo {author} {\bibfnamefont {C.}~\bibnamefont {Silberhorn}},\
  }\href@noop {} {\bibfield  {journal} {\bibinfo  {journal} {Phys. Rev. Lett.}\
  }\textbf {\bibinfo {volume} {106}},\ \bibinfo {pages} {013603} (\bibinfo
  {year} {2011})}\BibitemShut {NoStop}%
\bibitem [{\citenamefont {Mosley}\ \emph {et~al.}(2008)\citenamefont {Mosley},
  \citenamefont {Lundeen}, \citenamefont {Smith}, \citenamefont {Wasylczyk},
  \citenamefont {U'Ren}, \citenamefont {Silberhorn},\ and\ \citenamefont
  {Walmsley}}]{Mosley2008a}%
  \BibitemOpen
  \bibfield  {author} {\bibinfo {author} {\bibfnamefont {P.~J.}\ \bibnamefont
  {Mosley}}, \bibinfo {author} {\bibfnamefont {J.~S.}\ \bibnamefont {Lundeen}},
  \bibinfo {author} {\bibfnamefont {B.~J.}\ \bibnamefont {Smith}}, \bibinfo
  {author} {\bibfnamefont {P.}~\bibnamefont {Wasylczyk}}, \bibinfo {author}
  {\bibfnamefont {A.~B.}\ \bibnamefont {U'Ren}}, \bibinfo {author}
  {\bibfnamefont {C.}~\bibnamefont {Silberhorn}}, \ and\ \bibinfo {author}
  {\bibfnamefont {I.~A.}\ \bibnamefont {Walmsley}},\ }\href {\doibase
  10.1103/PhysRevLett.100.133601} {\bibfield  {journal} {\bibinfo  {journal}
  {Phys. Rev. Lett.}\ }\textbf {\bibinfo {volume} {100}},\ \bibinfo {pages}
  {133601} (\bibinfo {year} {2008})}\BibitemShut {NoStop}%
\bibitem [{\citenamefont {Jin}\ \emph {et~al.}(2011)\citenamefont {Jin},
  \citenamefont {Zhang}, \citenamefont {Shimizu}, \citenamefont {Matsuda},
  \citenamefont {Mitsumori}, \citenamefont {Kosaka},\ and\ \citenamefont
  {Edamatsu}}]{Jin2011}%
  \BibitemOpen
  \bibfield  {author} {\bibinfo {author} {\bibfnamefont {R.-B.}\ \bibnamefont
  {Jin}}, \bibinfo {author} {\bibfnamefont {J.}~\bibnamefont {Zhang}}, \bibinfo
  {author} {\bibfnamefont {R.}~\bibnamefont {Shimizu}}, \bibinfo {author}
  {\bibfnamefont {N.}~\bibnamefont {Matsuda}}, \bibinfo {author} {\bibfnamefont
  {Y.}~\bibnamefont {Mitsumori}}, \bibinfo {author} {\bibfnamefont
  {H.}~\bibnamefont {Kosaka}}, \ and\ \bibinfo {author} {\bibfnamefont
  {K.}~\bibnamefont {Edamatsu}},\ }\href {\doibase 10.1103/PhysRevA.83.031805}
  {\bibfield  {journal} {\bibinfo  {journal} {Phys. Rev. A}\ }\textbf {\bibinfo
  {volume} {83}},\ \bibinfo {pages} {031805} (\bibinfo {year}
  {2011})}\BibitemShut {NoStop}%
\bibitem [{\citenamefont {Jin}\ \emph {et~al.}(2013{\natexlab{c}})\citenamefont
  {Jin}, \citenamefont {Wakui}, \citenamefont {Shimizu}, \citenamefont
  {Benichi}, \citenamefont {Miki}, \citenamefont {Yamashita}, \citenamefont
  {Terai}, \citenamefont {Wang}, \citenamefont {Fujiwara},\ and\ \citenamefont
  {Sasaki}}]{Jin2013PRA}%
  \BibitemOpen
  \bibfield  {author} {\bibinfo {author} {\bibfnamefont {R.-B.}\ \bibnamefont
  {Jin}}, \bibinfo {author} {\bibfnamefont {K.}~\bibnamefont {Wakui}}, \bibinfo
  {author} {\bibfnamefont {R.}~\bibnamefont {Shimizu}}, \bibinfo {author}
  {\bibfnamefont {H.}~\bibnamefont {Benichi}}, \bibinfo {author} {\bibfnamefont
  {S.}~\bibnamefont {Miki}}, \bibinfo {author} {\bibfnamefont {T.}~\bibnamefont
  {Yamashita}}, \bibinfo {author} {\bibfnamefont {H.}~\bibnamefont {Terai}},
  \bibinfo {author} {\bibfnamefont {Z.}~\bibnamefont {Wang}}, \bibinfo {author}
  {\bibfnamefont {M.}~\bibnamefont {Fujiwara}}, \ and\ \bibinfo {author}
  {\bibfnamefont {M.}~\bibnamefont {Sasaki}},\ }\href@noop {} {\bibfield
  {journal} {\bibinfo  {journal} {Phys. Rev. A}\ }\textbf {\bibinfo {volume}
  {87}},\ \bibinfo {pages} {063801} (\bibinfo {year}
  {2013}{\natexlab{c}})}\BibitemShut {NoStop}%
\bibitem [{\citenamefont {Miki}\ \emph {et~al.}(2007)\citenamefont {Miki},
  \citenamefont {Fujiwara}, \citenamefont {Sasaki},\ and\ \citenamefont
  {Wang}}]{Miki2007}%
  \BibitemOpen
  \bibfield  {author} {\bibinfo {author} {\bibfnamefont {S.}~\bibnamefont
  {Miki}}, \bibinfo {author} {\bibfnamefont {M.}~\bibnamefont {Fujiwara}},
  \bibinfo {author} {\bibfnamefont {M.}~\bibnamefont {Sasaki}}, \ and\ \bibinfo
  {author} {\bibfnamefont {Z.}~\bibnamefont {Wang}},\ }\bibfield  {booktitle}
  {\emph {\bibinfo {booktitle} {Applied Superconductivity, IEEE Transactions
  on}},\ }\href@noop {} {\bibfield  {journal} {\bibinfo  {journal} {IEEE Trans.
  Appl. Superconduct.}\ }\textbf {\bibinfo {volume} {17}},\ \bibinfo {pages}
  {285} (\bibinfo {year} {2007})}\BibitemShut {NoStop}%
\bibitem [{\citenamefont {Broome}\ \emph {et~al.}(2011)\citenamefont {Broome},
  \citenamefont {Almeida}, \citenamefont {Fedrizzi},\ and\ \citenamefont
  {White}}]{Broome2011}%
  \BibitemOpen
  \bibfield  {author} {\bibinfo {author} {\bibfnamefont {M.~A.}\ \bibnamefont
  {Broome}}, \bibinfo {author} {\bibfnamefont {M.~P.}\ \bibnamefont {Almeida}},
  \bibinfo {author} {\bibfnamefont {A.}~\bibnamefont {Fedrizzi}}, \ and\
  \bibinfo {author} {\bibfnamefont {A.~G.}\ \bibnamefont {White}},\ }\bibfield
  {booktitle} {\emph {\bibinfo {booktitle} {Optics Express}},\ }\href@noop {}
  {\bibfield  {journal} {\bibinfo  {journal} {Opt. Express}\ }\textbf {\bibinfo
  {volume} {19}},\ \bibinfo {pages} {22698} (\bibinfo {year}
  {2011})}\BibitemShut {NoStop}%
\bibitem [{\citenamefont {Hong}\ \emph {et~al.}(1987)\citenamefont {Hong},
  \citenamefont {Ou},\ and\ \citenamefont {Mandel}}]{Hong1987}%
  \BibitemOpen
  \bibfield  {author} {\bibinfo {author} {\bibfnamefont {C.~K.}\ \bibnamefont
  {Hong}}, \bibinfo {author} {\bibfnamefont {Z.~Y.}\ \bibnamefont {Ou}}, \ and\
  \bibinfo {author} {\bibfnamefont {L.}~\bibnamefont {Mandel}},\ }\href@noop {}
  {\bibfield  {journal} {\bibinfo  {journal} {Phys. Rev. Lett.}\ }\textbf
  {\bibinfo {volume} {59}},\ \bibinfo {pages} {2044} (\bibinfo {year}
  {1987})}\BibitemShut {NoStop}%
\bibitem [{\citenamefont {Kuzucu}\ \emph {et~al.}(2005)\citenamefont {Kuzucu},
  \citenamefont {Fiorentino}, \citenamefont {Albota}, \citenamefont {Wong},\
  and\ \citenamefont {K{\"a}rtner}}]{Kuzucu2005}%
  \BibitemOpen
  \bibfield  {author} {\bibinfo {author} {\bibfnamefont {O.}~\bibnamefont
  {Kuzucu}}, \bibinfo {author} {\bibfnamefont {M.}~\bibnamefont {Fiorentino}},
  \bibinfo {author} {\bibfnamefont {M.~A.}\ \bibnamefont {Albota}}, \bibinfo
  {author} {\bibfnamefont {F.~N.~C.}\ \bibnamefont {Wong}}, \ and\ \bibinfo
  {author} {\bibfnamefont {F.~X.}\ \bibnamefont {K{\"a}rtner}},\ }\href@noop {}
  {\bibfield  {journal} {\bibinfo  {journal} {Phys. Rev. Lett.}\ }\textbf
  {\bibinfo {volume} {94}},\ \bibinfo {pages} {083601} (\bibinfo {year}
  {2005})}\BibitemShut {NoStop}%
\bibitem [{\citenamefont {Shimizu}\ and\ \citenamefont
  {Edamatsu}(2009)}]{Shimizu2009}%
  \BibitemOpen
  \bibfield  {author} {\bibinfo {author} {\bibfnamefont {R.}~\bibnamefont
  {Shimizu}}\ and\ \bibinfo {author} {\bibfnamefont {K.}~\bibnamefont
  {Edamatsu}},\ }\bibfield  {booktitle} {\emph {\bibinfo {booktitle} {Optics
  Express}},\ }\href@noop {} {\bibfield  {journal} {\bibinfo  {journal} {Opt.
  Express}\ }\textbf {\bibinfo {volume} {17}},\ \bibinfo {pages} {16385}
  (\bibinfo {year} {2009})}\BibitemShut {NoStop}%
\bibitem [{\citenamefont {Ansari}\ \emph {et~al.}(2014)\citenamefont {Ansari},
  \citenamefont {Brecht}, \citenamefont {Harder},\ and\ \citenamefont
  {Silberhorn}}]{Ansari2014}%
  \BibitemOpen
  \bibfield  {author} {\bibinfo {author} {\bibfnamefont {V.}~\bibnamefont
  {Ansari}}, \bibinfo {author} {\bibfnamefont {B.}~\bibnamefont {Brecht}},
  \bibinfo {author} {\bibfnamefont {G.}~\bibnamefont {Harder}}, \ and\ \bibinfo
  {author} {\bibfnamefont {C.}~\bibnamefont {Silberhorn}},\ }\href@noop {}
  {\bibfield  {journal} {\bibinfo  {journal} {arXiv:1404.7725}\ } (\bibinfo
  {year} {2014})}\BibitemShut {NoStop}%
\bibitem [{\citenamefont {Ma}\ \emph {et~al.}(2011)\citenamefont {Ma},
  \citenamefont {Zotter}, \citenamefont {Kofler}, \citenamefont {Jennewein},\
  and\ \citenamefont {Zeilinger}}]{Ma2011}%
  \BibitemOpen
  \bibfield  {author} {\bibinfo {author} {\bibfnamefont {X.-S.}\ \bibnamefont
  {Ma}}, \bibinfo {author} {\bibfnamefont {S.}~\bibnamefont {Zotter}}, \bibinfo
  {author} {\bibfnamefont {J.}~\bibnamefont {Kofler}}, \bibinfo {author}
  {\bibfnamefont {T.}~\bibnamefont {Jennewein}}, \ and\ \bibinfo {author}
  {\bibfnamefont {A.}~\bibnamefont {Zeilinger}},\ }\href@noop {} {\bibfield
  {journal} {\bibinfo  {journal} {Phys. Rev. A}\ }\textbf {\bibinfo {volume}
  {83}},\ \bibinfo {pages} {043814} (\bibinfo {year} {2011})}\BibitemShut
  {NoStop}%
\bibitem [{\citenamefont {Morris}\ \emph {et~al.}(2014)\citenamefont {Morris},
  \citenamefont {Francis-Jones}, \citenamefont {Wilcox}, \citenamefont
  {Tropper},\ and\ \citenamefont {Mosley}}]{Morris2014}%
  \BibitemOpen
  \bibfield  {author} {\bibinfo {author} {\bibfnamefont {O.~J.}\ \bibnamefont
  {Morris}}, \bibinfo {author} {\bibfnamefont {R.~J.}\ \bibnamefont
  {Francis-Jones}}, \bibinfo {author} {\bibfnamefont {K.~G.}\ \bibnamefont
  {Wilcox}}, \bibinfo {author} {\bibfnamefont {A.~C.}\ \bibnamefont {Tropper}},
  \ and\ \bibinfo {author} {\bibfnamefont {P.~J.}\ \bibnamefont {Mosley}},\
  }\bibfield  {booktitle} {\emph {\bibinfo {booktitle} {Special Issue on
  Nonlinear Quantum Photonics}},\ }\href@noop {} {\bibfield  {journal}
  {\bibinfo  {journal} {Opt. Commun.}\ }\textbf {\bibinfo {volume} {327}},\
  \bibinfo {pages} {39} (\bibinfo {year} {2014})}\BibitemShut {NoStop}%
\bibitem [{Gig()}]{GigaOptics}%
  \BibitemOpen
  \href@noop {} {\enquote {\bibinfo {title} {Giga optics website,
  http://www.gigaoptics.com/},}\ }\BibitemShut {NoStop}%
\bibitem [{M2l()}]{M2laser}%
  \BibitemOpen
  \href@noop {} {\enquote {\bibinfo {title} {M2 laser website,
  http://www.m2lasers.com/},}\ }\BibitemShut {NoStop}%
\end{thebibliography}
%


\section*{Appendix-I}

\begin{figure}[htbp]
\includegraphics[width=0.45 \textwidth]{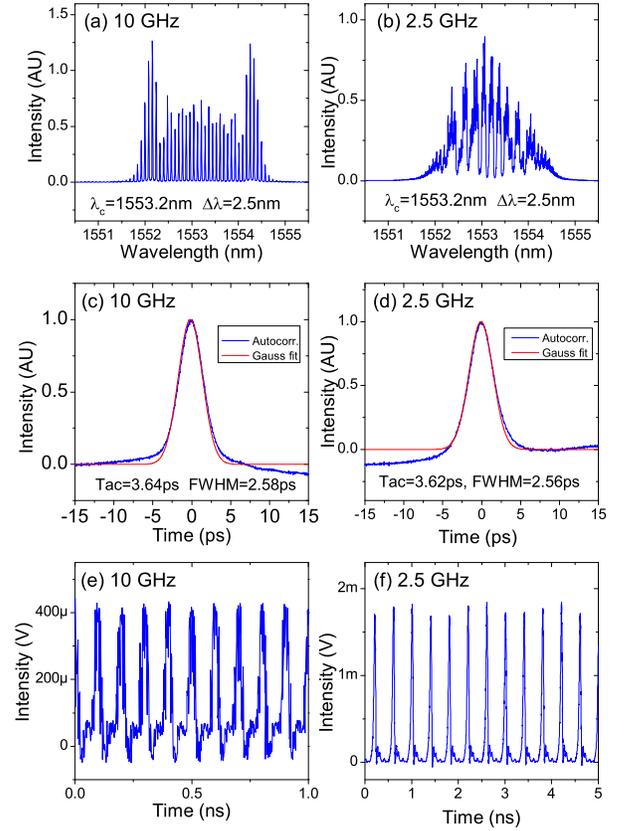}
\caption{(a-f) The spectra, autocorrelation, and temporal sequences of the comb laser at 10 GHz and 2.5 GHz repetition rates.
The full-width-at-half-maximum (FWHM) of the autocorrelation data are round 3.6 ps, corresponding to FWHM of 2.6 ps for the fundamental lasers.}
\label{otherspectra}
\end{figure}

\begin{figure}[htbp]
\includegraphics[width=0.35 \textwidth]{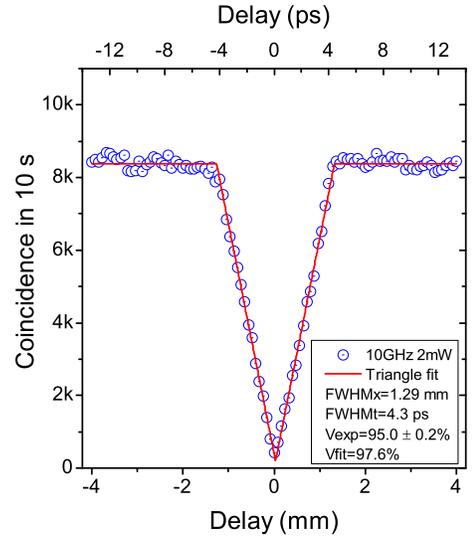}
\caption{Hong-Ou-Mandel dip for comb laser at 10 GHz repetition rate, fitted with triangle shape function. The pump power is 2 mW. }
 \label{otherHOMI}
\end{figure}

In this part we provide more information of the  comb laser at 10 GHz and 2.5 GHz repetition rates.
Figure\,\ref{otherspectra} compares the spectra, autocorrelation, and temporal sequences of the comb laser at 10 GHz and 2.5 GHz repetition rates.
Figure\,\ref{otherHOMI} shows the Hong-Ou-Mandel dip for the comb laser at 10 GHz, with a similar bandwidth and visibility as the results by the laser at 2.5 GHz repetition rate.

\section*{Appendix-II}

In this part, we numerically analyze the relationship between photon-pair generation rate (i.e., average photon pair per pulse) and HOM interference visibility.

\subsection{The model}
\begin{figure}
\begin{center}
\includegraphics[width=70mm]{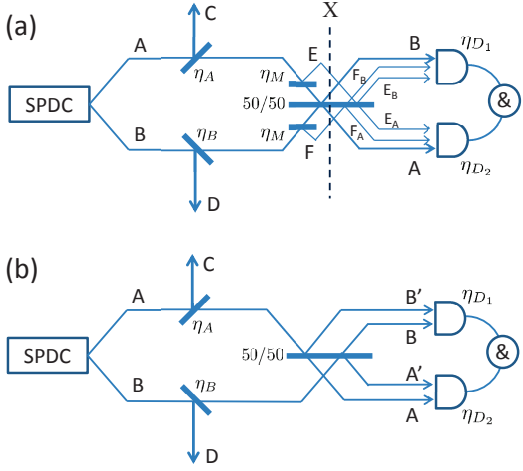}   %
\caption{\label{fig:model}
Models. (a) No delay (HOM dip). The mode mismatch at the 50/50 beam splitter
is represented by the beam splitters $\eta_M$.
(b) With delay. The delay is represnted by the spatial difference
at the 50/50 beam splitter.
}
\end{center}
\end{figure}

Here, we describe a numerical model of the HOM experiment.
The model is described in Fig.~\ref{fig:model}(a) (without delay) and (b)
(with delay) where $\eta_{A,B}$ represent
transmittances of mode $A$ and $B$ (losses are effectively described by
beam splitters), respectively, and $\eta_{D_1}$ and $\eta_{D_2}$ are
the detector efficiencies.
The mode mismatch between the signal and idler pulses
is directly reflected to the HOM interference visibility.
In our model, the mode matching efficiency $\eta_M$ is effectively represented by two  beam splitters with the transmittance $\eta_M$  and extra modes $E$ and $F$.

The HOM interference visibility is defined as
\begin{equation}
\label{eq:HOM_def}
V = \frac{CC_{\rm mean} - CC_{\rm min}}{CC_{\rm mean}} ,
\end{equation}
where $CC_{\rm min}$ and $CC_{\rm mean}$ are the coincidence count rates
with zero-delay and large delay (i.e., with and without interference
between the signal and idler), respectively.
In the following we derive $CC_{\rm min}$ and $CC_{\rm mean}$ separately
from our model.

The initial state from the SPDC source is given by
a two-mode squeezed-vacuum state
\begin{equation}
\label{eq:TMSV}
|\psi\rangle_{AB} = \sqrt{1-\lambda^2} \sum_{n=0}^\infty \lambda^n
|n\rangle_A |n\rangle_B ,
\end{equation}
where $\lambda$ is the squeezing parameter, and $\lambda^2/(1-\lambda^2) = p $ is the average photon pairs per pulse.
Let $\hat{V}_{A(B)}^\eta$ be a beam splitting operator on mode $A(B)$
with transmittance $\eta$ which transforms the photon number states
$|n_1\rangle|n_2\rangle$ as
\begin{eqnarray}
\label{eq:BS}
\hat{V}_{AB}^\eta |n_1\rangle|n_2\rangle
& = & \frac{1}{\sqrt{n_1! n_2!}} \sum_{k_1=0}^{n_1} \sum_{k_2=0}^{n_2}
\binom{n_1}{k_1} \binom{n_2}{k_2} (-1)^{k_2}
\nonumber\\ && \times
\sqrt{\eta}^{n_2+k_1-k_2} \sqrt{1-\eta}^{n_1-k_1+k_2}
\nonumber\\ && \times
\sqrt {(k_1  + k_2 )!(n_1  + n_2  - k_1  - k_2 )!}
\nonumber\\ && \times
\left| {k_1  + k_2 } \right\rangle \left| {n_1  + n_2  - k_1  - k} \right\rangle .
\end{eqnarray}
Applying the beam splitters $\hat{V}^{\eta_A}_{AC}$, $\hat{V}^{\eta_B}_{BD}$,
$\hat{V}^{\eta_M}_{AE}$, $\hat{V}^{\eta_M}_{BF}$, and $\hat{V}^{1/2}_{AB}$
onto the two-mode squeezed vacuum (state at X in Fig.~\ref{fig:model}(a)),
we obtain
\begin{widetext}
\begin{eqnarray}
\label{eq:BS_operations}
&&
\hat{V}^{1/2}_{AB} \hat{V}^{\eta_M}_{BF} \hat{V}^{\eta_M}_{AE}
\hat{V}^{\eta_B}_{BD} \hat{V}^{\eta_A}_{AC}
|\psi\rangle_{AB} |0\rangle_C |0\rangle_D |0\rangle_E |0\rangle_F
\nonumber\\
&&
= \sqrt{1-\lambda^2} \sum_{n=0}^\infty \lambda^n
\sum_{k_1=0}^n \binom{n}{k_1}^{1/2}
\eta_A^{k_1/2} (1-\eta_A)^{\frac{n-k_1}{2}}
\sum_{k_2=0}^n \binom{n}{k_2}^{1/2}
\eta_B^{k_2/2} (1-\eta_B)^{\frac{n-k_2}{2}}
\nonumber\\ && \quad \times
\sum_{k_3=0}^{k_1} \binom{k_1}{k_3}^{1/2}
\eta_M^{k_3/2} (1-\eta_M)^{\frac{k_1-k_3}{2}}
\sum_{k_4=0}^{k_2} \binom{k_2}{k_4}^{1/2}
\eta_M^{k_4/2} (1-\eta_M)^{\frac{k_2-k_4}{2}}
\nonumber\\ && \quad \times
\sum_{k_5=0}^{k_3} \sum_{k_6=0}^{k_4} \binom{k_3}{k_5} \binom{k_4}{k_6}
\left(\frac{1}{2}\right)^{\frac{k_3+k_4}{2}} (-1)^{k_6}
\left( \frac{(k_5 + k_6)!\, (k_3+k_4-k_5-k_6)!}{k_3!\,k_4!} \right)^{1/2}
\nonumber\\ && \quad \times
|k_5+k_6\rangle_A |k_3+k_4-k_5-k_6\rangle_B |n-k_1\rangle_C |n-k_2\rangle_D
|k_1-k_3\rangle_E |k_2-k_4\rangle_F
\nonumber\\
&&
= \sqrt{1-\lambda^2} \sum_{n=0}^\infty \lambda^n
\sum_{k_1=0}^n \binom{n}{k_1}^{1/2}
\eta_A^{k_1/2} (1-\eta_A)^{\frac{n-k_1}{2}}
\sum_{k_2=0}^n \binom{n}{k_2}^{1/2}
\eta_B^{k_2/2} (1-\eta_B)^{\frac{n-k_2}{2}}
\nonumber\\ && \quad \times
\sum_{k_3=0}^{k_1} \binom{k_1}{k_3}^{1/2}
\eta_M^{k_3/2} (1-\eta_M)^{\frac{k_1-k_3}{2}}
\sum_{k_4=0}^{k_2} \binom{k_2}{k_4}^{1/2}
\eta_M^{k_4/2} (1-\eta_M)^{\frac{k_2-k_4}{2}}
\left( \frac{1}{2} \right)^{k_3+k_4}
\nonumber\\ && \quad \times
\sum_{l=0}^{k_3 + k_4} \sum_{k_5=\max\{0, l-k_4\}}^{\min\{l,k_3\}}
(-1)^{l-k_5} \left\{
\binom{k_3}{k_5} \binom{k_4}{l-k_5} \binom{l}{k_5} \binom{k_3+k_4-l}{k_3-k_5}
\right\}^{1/2}
\nonumber\\ && \quad \times
|l\rangle_A |k_3+k_4-l\rangle_B |n-k_1\rangle_C |n-k_2\rangle_D
|k_1-k_3\rangle_E |k_2-k_4\rangle_F ,
\end{eqnarray}
where $l=k_5 + k_6$ and we have used the relation
\begin{equation}
\label{eq:binom}
\binom{k_3}{k_5} \binom{k4}{k_6}
\left( \frac{(k_5 + k_6)!\, (k_3+k_4-k_5-k_6)!}{k_3!\,k_4!} \right)^{1/2}
= \left\{ \binom{k_3}{k_5} \binom{k4}{k_6} \binom{k_5+k_6}{k_5}
\binom{k_3+k_4-k_5-k_6}{k_3-k_5} \right\}^{1/2} .
\end{equation}
\end{widetext}
Note that $\hat{V}^{1/2}$ should be applied to mode $E$ and $F$,
which will be discussed later.
From Eq.~(\ref{eq:BS_operations}) we find the joint probability
of having $l$, $k_3+k_4-l$, $n-k_1$, $n-k_2$, $k_1-k_3$, $k_2-k_4$ photons
in mode A-F at X:
\begin{widetext}
\begin{eqnarray}
\label{eq:joint_prob_X}
&& P^X_{ABCDEF}(l, k_3+k_4-l, n-k_1, n-k_2, k_1-k_3, k_2-k_4)
\nonumber\\ &&
= (1-\lambda)^2 \lambda^{2n}
\eta_A^{k_1} (1-\eta_A)^{n-k_1} \eta_B^{k_2} (1-\eta_B)^{n-k_2}
\eta_M^{k_3+k_4} (1-\eta_M)^{k_1+k_2-k_3-k_4}
\left( \frac{1}{2} \right)^{k_3+k_4}
\binom{n}{k_1} \binom{n}{k_2} \binom{k_1}{k_3} \binom{k_2}{k_4}
\nonumber\\ && \quad \times
\left\{
\sum_{k_5=\max\{0, l-k_4\}}^{\min\{l,k_3\}} (-1)^{l-k_5} \left\{
\binom{k_3}{k_5} \binom{k_4}{l-k_5} \binom{l}{k_5} \binom{k_3+k_4-l}{k_3-k_5}
\right\}^{1/2}
\right\}^2 .
\end{eqnarray}
\end{widetext}
The 50/50 beam splitting of mode $E$ ($F$) into $E_A$ and $E_B$
($F_A$ and $F_B$) adds extra binomial distribution terms
$\binom{k_1-k_3}{k_7} \binom{k_2-k_4}{k_8}
\left( \frac{1}{2} \right)^{k_1+k_2-k_3-k_4}$ to
Eq.~(\ref{eq:joint_prob_X}). The joint probability distribution
for the state at the detectors is thus given by
\begin{widetext}
\begin{eqnarray}
\label{eq:joint_prob}
&& P_{ABCDE_AF_AE_BF_B}(l, k_3+k_4-l, n-k_1, n-k_2, k_7, k_2-k_4-k_8,
k_1-k_3-k_7, k_8)
\nonumber\\ &&
= (1-\lambda)^2 \lambda^{2n}
\eta_A^{k_1} (1-\eta_A)^{n-k_1} \eta_B^{k_2} (1-\eta_B)^{n-k_2}
\eta_M^{k_3+k_4} (1-\eta_M)^{k_1+k_2-k_3-k_4}
\left( \frac{1}{2} \right)^{k_1+k_2}
\binom{n}{k_1} \binom{n}{k_2} \binom{k_1}{k_3} \binom{k_2}{k_4}
\nonumber\\ && \quad \times
\binom{k_1-k_3}{k_7} \binom{k_2-k_4}{k_8}
\left\{
\sum_{k_5=\max\{0, l-k_4\}}^{\min\{l,k_3\}} (-1)^{l-k_5} \left\{
\binom{k_3}{k_5} \binom{k_4}{l-k_5} \binom{l}{k_5} \binom{k_3+k_4-l}{k_3-k_5}
\right\}^{1/2}
\right\}^2 .
\end{eqnarray}
The coincidence rate $CC_{\rm min}$ is then obtained by the sum of
the joint probability:
\begin{eqnarray}
\label{eq:C_min}
CC_{\rm min} & = & \sum_{n=0}^\infty \sum_{k_1=0}^n \sum_{k_2=0}^n
\sum_{k_3=0}^{k_1} \sum_{k_4=0}^{k_2}
\sum_{k_7=0}^{k_1-k_3} \sum_{k_8=0}^{k_2-k_4} \sum_{l=0}^{k_3+k_4}
\left\{ 1- (1-\eta_{D_1})^{l+k_2-k_4+k_7-k_8} \right\}
\left\{ 1- (1-\eta_{D_2})^{-l+k_1+k_4-k_7+k_8} \right\}
\nonumber\\ &&
\times P_{ABCDE_AF_AE_BF_B}(l, k_3+k_4-l, n-k_1, n-k_2, k_7, k_2-k_4-k_8,
k_1-k_3-k_7, k_8) .
\end{eqnarray}
\end{widetext}

The derivation of $CC_{\rm mean}$ is rather simple since there is no
interference at the 50/50 beam splitter due to the delay.
This is illustrated in Fig.~\ref{fig:model}(b).
Note that we do not need $\eta_M$.
The two-mode squeezed vacuum from the SPDC source has
a joint photon distribution:
\begin{equation}
\label{eq:P_AB}
P_{AB}(n,n) = (1-\lambda^2) \lambda^{2 n} .
\end{equation}
The beam splitting operation simply spread this distribution
in a binomial manner. For example, after the beam splitter $\eta_A$,
the joint distribution is given by
\begin{equation}
\label{eq:P_AB}
P_{ABC}(n, k_1, n-k_1) = (1-\lambda^2) \lambda^{2 n} \binom{n}{k_1}
\eta_A^{k_1} (1-\eta_A)^{n-k_1}.
\end{equation}
Applying the $\eta_B$ and 50/50 beam splitters in a similar way, we have
\begin{widetext}
\begin{eqnarray}
\label{eq:P_AA'BB'CD}
&& P_{AA'BB'CD}(k_3, k_2-k_4, k_4, k_1-k_3, n-k_1, n-k_2)
\nonumber\\ &&
= (1-\lambda^2) \lambda^{2 n} \binom{n}{k_1} \binom{n}{k_2}
\binom{k_1}{k_3} \binom{k_2}{k_4} \eta_A^{k_1} (1-\eta_A)^{n-k_1}
\eta_B^{k_2} (1-\eta_B)^{n-k_2}
\left( \frac{1}{2} \right)^{k_1+k_2} ,
\end{eqnarray}
before the detectors.
The coincidence count $CC_{\rm mean}$ is then given by
\begin{eqnarray}
\label{eq:C_mean}
CC_{\rm mean} & = & \sum_{n=0}^\infty \sum_{k_1=0}^n \sum_{k_2=0}^n
\sum_{k_3=0}^{k_1} \sum_{k_4=0}^{k_2}
\left\{ 1- (1-\eta_{D_1})^{k_2+k_3-k_4} \right\}
\left\{ 1- (1-\eta_{D_2})^{k_1-k_3+k_4} \right\}
\nonumber\\ &&
\times P_{AA'BB'CD}(k_3, k_2-k_4, k_4, k_1-k_3, n-k_1, n-k_2) .
\end{eqnarray}
\end{widetext}
The HOM visibility in Eq.~(\ref{eq:HOM_def}) is thus calculable from
Eqs.~(\ref{eq:C_min}) and (\ref{eq:C_mean}).

\subsection{Numerical result}

The transmittances (efficiencies) of each components in the experiment
are summarized in Table~\ref{table:parameters}
(see Fig.~\ref{fig:model} for the theoretical model and
the corresponding experimental setup in Main text.
In fact, the HOM visibility is extremely sensitive to the mode matching
factor $\eta_M$. It is however not easy to estimate
the mode matching factor $\eta_M$ experimentally with enough accuracy.

\begin{figure}[htbp]
\begin{center}
\includegraphics[width=70mm]{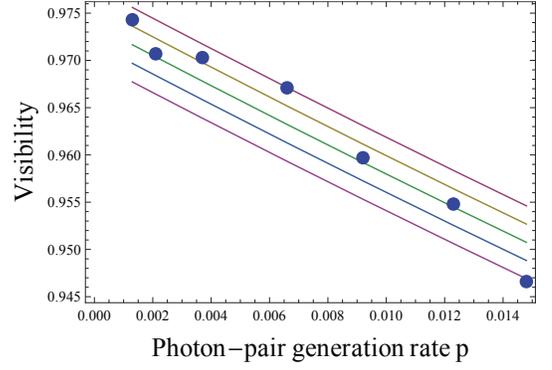}   %
\caption{\label{fig:numerics}
The HOM visibility versus $p$.
The solid lines represent theoretical curves with
$\eta_M=0.9888, 0.9878, 0.9868, 0.9858, 0.9848$
from the top to the bottom. The plots are the experimental results
with the 76 MHz laser.
}
\end{center}
\end{figure}

\begin{figure}[H]
\begin{center}
\includegraphics[width=70mm]{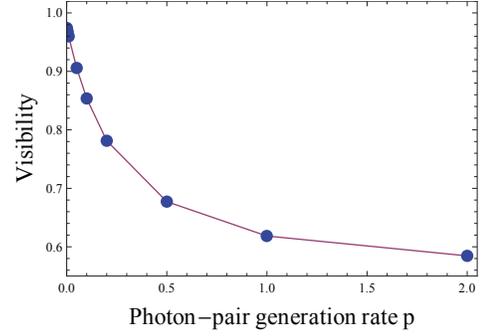}   %
\caption{\label{fig:nVSv}
The HOM visibilities at different $p$ values, with  $\eta_M=0.9878$.
}
\end{center}
\end{figure}

In Fig.~\ref{fig:numerics}, we plot the numerical results with
various $\eta_M$, and the experimental data with the 76 MHz laser.
The experimental average photon-pair $p$ is estimated from
the experimental count rates.
The experimental data fit the theoretical lines well
within $0.9848 \le \eta_M \le 0.9888$.
With the parameters in  Table~\ref{table:parameters}, we also calculated the performance of our scheme at high   photon-pair generation rate, as shown in  Fig.~\ref{fig:nVSv} and  Table~\ref{table:nVSv}.
From this simulation, we find several interesting relationship.
(1), The visibility is directly determined by the average photon-pairs.
(2), The slope of this line is very sensitive to the unbalanced loss in the delay arm and non-delay arm.
(3), The Y-intercept of this line very sensitive  to the mode matching efficiency.

\begin{table}
\caption{\label{table:parameters} Transmittance and efficiency of the components in the experiment.
SMFC: single mode fiber coupler.
FC: fiber connector. SNSPD: superconducting nanowier single photon detector.}
\begin{tabular}{cll}
\hline \hline
$\eta_A$ \qquad \qquad & 0.42 \qquad \qquad & SMFC + FCs \\
$\eta_B$ \qquad \qquad & 0.29 \qquad \qquad & SMFC + FCs + Delay line \\
$\eta_{D_1}$ \qquad \qquad & 0.68 \qquad \qquad & SNSPD1 \\
$\eta_{D_2}$ \qquad \qquad & 0.70 \qquad \qquad & SNSPD2 \\
\hline \hline
\end{tabular}
\end{table}
%
%
%
%
%
%
%
%
%
%
%
%
\begin{table}[H]
\caption{\label{table:nVSv}  The visibilities at different photon-pair generation rate.}
\begin{tabular}{clllllllllll}
\hline \hline
$p$              &0.001   &0.005  &0.01    &0.05   &0.1      &0.2     &0.5     &1           &2        \\
V               &0.974   &0.968  &0.960   &0.906  &0.854    &0.781   &0.677   &0.618       &0.585    \\
\hline \hline
\end{tabular}
\end{table}

\end{document}